\documentclass[nofootinbib,11pt]{revtex4}
\usepackage{amssymb}

\usepackage{epsfig}

\begin{document}

\title{On the possibility of metamaterial properties in spin plasmas}
\author{G. Brodin and M. Marklund\footnote{E-mail: mattias.marklund@physics.umu.se}}

\affiliation{Department of Physics, Ume{\aa} University, SE--901 87
Ume{\aa}, Sweden}

\begin{abstract}
The fluid theory of plasmas is extended to include the properties of
electron spin. The linear theory of waves in a magnetized plasma is
presented, and it is shown that the spin effects causes a change of the
magnetic permeability. Furthemore, by changing the direction of the external
magnetic field, the magnetic permability may become negative. This leads to
instabilities in the long wavelength regimes. If these can be controlled,
however, the spin plasma becomes a metamaterial for a broad range of
frequencies, i.e. above the ion cyclotron frequency but below the electron
cyclotron frequency. The consequences of our results are discussed.
\end{abstract}

\maketitle

\section{Introduction}

The theoretical concept of materials with a negative index of refraction has a long history,
while the practical developments are relatively recent in this field.  
In his Moscow school, Mandelstam presented results for
materials exhibiting a negative index of refraction \cite
{Mandelstam1,Mandelstam2} (see also the historical account in Ref.\ \cite
{Agranovich-Gartstein}). Negative permittivity
and permeability materials, and the consequences of a negative group velocity, were
moreover given attention by Pafomov \cite{Pafomov} and Agranovich and
Ginzburg \cite{Agranovich-Ginzburg}, and have since then been discussed by
several authors, most notably by Veselago \cite{Veselago}, who gave a detailed analysis
of the consequences of such material properties (see also \cite{Agranovich1,Agranovich2,Zhang,Agranovich-Gartstein}). 
Although not known to be found naturally, such materials have recently been realized in laboratory environments \cite{Smith-etal,Shelby-etal},
and the experimental development in conjunction with theoretical insights (see Refs. \cite{Leonhardt,Pendry-etal}) has spawned a rapidly growing interest in these materials (see, e.g. \cite{Pendry,Ramakrishna} for a review).
From a sharp resonance in the material response to the applied external
field, one may obtain negative $\varepsilon$ and $\mu$. The normal procedure
for obtaining negative-index-of-refraction materials is to put together two
structured materials that both have negative permittivity and negative
permeability, such that the resulting composite material has a negative
refractive index \cite{Ramakrishna,Krowne-Zhang}. A nonlinear metamaterial can also be constructed through the nonlinear properties of the constituent materials \cite
{DAguanno-etal,Scalora-etal}, admitting new types of solitary wave structures
\cite{Marklund-Shukla-Stenflo-Brodin,Marklund-Shukla-Stenflo,Shadrivov-Kivshar}.

The field of quantum plasmas is a rapidly growing field of research. From
the non-relativistic domain, with its basic description in terms of the
Schr\"odinger equation, to the strongly relativistic regime, with its
natural connection to quantum field theory, quantum plasma physics provides
promises of highly interesting and important application, fundamental
connections between different areas of science, as well as difficult
challenges from a computational perspective. The necessity to thoroughly
understand such plasmas motivates a reductive principle of research, for
which we successively build more complex models based on previous results.
The simplest lower order effect due to relativistic quantum mechanics is the
introduction of spin, and as such thus provides a first step towards a
partial description of relativistic quantum plasmas.

Already in the 1960's, Pines studied the excitation spectrum of quantum
plasmas \cite{Pines,pines-book}, for which we have a high density and a low
temperature as compared to normal plasmas.  Recently there has been an
increased interest in the properties of quantum plasmas \cite
{Manfredi2005,haas-etal1,haas,shukla,garcia-etal,collection2C,collection2E,collection2G,collection2H,collection2I,collection2K,collection2L,Haas-HarrisSheet,marklund-brodin,brodin-marklund,BM-pairplasma,shukla-eliasson,shukla-eliasson2,shaikh-shukla,brodin-marklund2}%
. The studies has been motivated by the development in nanostructured
materials \cite{craighead} and quantum wells \cite{manfredi-hervieux}, the
discovery of ultracold plasmas \cite{li-etal,fletcher-etal}, or a general
theoretical interest. The list of quantum mechanical effects that can be
included in a fluid picture includes the dispersive particle properties
accounted for by the Bohm potential \cite
{Manfredi2005,haas-etal1,haas,shukla,garcia-etal,collection2C,collection2E,collection2G,collection2H,collection2I,collection2K,collection2L,Haas-HarrisSheet}%
, the zero temperature Fermi pressure \cite
{Manfredi2005,haas-etal1,haas,shukla,garcia-etal}, spin properties \cite
{marklund-brodin,brodin-marklund,BM-pairplasma,brodin-marklund2,Brodin-Marklund-Manfredi}
as well as certain quantum electrodynamical effects \cite
{marklund-shukla,Lundin2007,lundstrom-etal,Brodin-etal-2007}. Within such
descriptions, \cite
{Manfredi2005,haas-etal1,haas,shukla,garcia-etal,marklund-brodin,brodin-marklund,Lundin2007,lundstrom-etal,Brodin-etal-2007}
quantum and classical collective effects can be described within a unified
picture, sometimes even showing a surprising overlap between classical and quantum behaviour 
\cite{Brodin-Marklund-Manfredi}.

Here we study the linear theory of electromagnetic wave propagation in a
magnetized plasma, with a special focus on the properties caused by the
electron spin. We are then able to present a scheme for
such a system to display metamaterial behavior. Specifically this is
induced by exposing a low temperature high density plasma to an external
magnetic field, which creates a magnetization in the plasma due to the
electron spin.  By changing the direction of the external magnetic field,
the magnetic permeability may become negative. It should be noted that the
above procedure induces instabilities in the long wavelength regime. A
number of ways to control these instabilities are pointed out. Assuming that
this can be done successfully, the spin plasma becomes a metamaterial for a
broad range of frequencies, i.e. above the ion cyclotron frequency but below
the electron cyclotron frequency. The conditions needed to create a
sufficient magnetization is discussed in the final section of the manuscript.

\section{Basic equations}

The theory for quantum plasmas including the effects of particle dispersion 
\cite
{Manfredi2005,haas-etal1,haas,shukla,garcia-etal,collection2C,collection2E,collection2G,collection2H,collection2I,collection2K,collection2L,Haas-HarrisSheet}%
, the Fermi pressure \cite{Manfredi2005,haas-etal1,haas,shukla,garcia-etal}
and effects due to the electron spin \cite
{marklund-brodin,brodin-marklund,BM-pairplasma} has been described in a
number of recent papers. For our purposes it will be sufficient to include
the spin effects, as the Fermi pressure and the Bohm-De Broigle potential
will not affect whether the plasma is a metamaterial or not.\footnote{%
This is clear from the dispersionrelation presented in [52], where the effect
of the Fermi pressure and the Bohm de Broigle potential is incorpatated in
an effective thermal velocity, as far as the linear wave modes are concerned.%
} Furthermore, neglecting terms that are quadratic in the spin vector,%
\footnote{%
The terms that are quadratic in the spin vector can be omitted if the
characteristic spatial scale is longer than the thermal de-Broigle
wavelength.} the governing spin plasma equations can be written 
\begin{equation}
\frac{\partial n_{s}}{\partial t}+\nabla \cdot \left( n_{s}\mathbf{v}%
_{s}\right) =0,  \label{Eq:cont}
\end{equation}

\begin{equation}
m_{s}n_{s}\left( \frac{\partial }{\partial t}+\mathbf{v}_{s}\cdot \nabla
\right) \mathbf{v}_{s}=q_{s}\left( \mathbf{E}+\mathbf{v}_{s}\times \mathbf{B}%
\right) +\frac{2\mu _{s}}{\hbar }s^{a}\nabla B_{a}-\nabla P_{s},
\label{Eq:momentum}
\end{equation}
together with the spin evolution equation 
\begin{equation}
\left( \frac{\partial }{\partial t}+\mathbf{v}_{s}\cdot \nabla \right) 
\mathbf{s=}\frac{2\mu _{s}}{\hbar }\mathbf{s_s\times B.}  \label{Eq:spin}
\end{equation}
Here $q_{s}$ and $m_{s}$ is the charge and mass of species $s$, $\mu
_{s}=q_{s}\hbar /2m_{s}$ is the magnetic moment, $P_{s}=$ $%
k_{B}T\nabla n_{s}$ is the pressure(for simplicity we use an
isothermal pressure model), $n_{s}$ is the number density, $\mathbf{v}_{s}$
is the velocity and $\mathbf{s}_s$ is the spin vector.

Next we concentrate on the linear wave modes in a magnetized plasma
described by Eqs. (\ref{Eq:cont})-(\ref{Eq:spin}) together with Maxwell's
equations 
\begin{equation}
\nabla \times \mathbf{E=-}\frac{\partial \mathbf{B}}{\partial t}
\label{Eq:Faraday}
\end{equation}
and 
\begin{equation}
\nabla \times \mathbf{B=}\mu _{0}\mathbf{J+}\frac{1}{c^{2}}\frac{\partial 
\mathbf{E}}{\partial t}  \label{Eq:Ampere}
\end{equation}
where, in addition to the free current density, we include the magnetization current source 
\begin{equation}
\mathbf{J}_{m}=\nabla \times \mathbf{M}_s = \nabla \times \left( \frac{2\mu
_{s}n_{s}}{\hbar }\mathbf{s}_s\right) ,  \label{Eq:magnetization}
\end{equation}
For this purpose we chose a coordinate system such that the unperturbed
magnetic field is $\mathbf{B}_{0}=B_{0}\widehat{\mathbf{z}}$, the wavevector
is $\mathbf{k}=k_{\bot }\widehat{\mathbf{x}}+k_{z}\widehat{\mathbf{z}}$, and
the variables are divided into an equilibrium value (index 0) and a
perturbed part (index 1). For simplicity index 1 are omitted on the electric
field and the velocity, since these variables have a zero unperturbed part.
As a preparation, we first consider the linear theory without the spin
terms. Using the continuity equation to express the density in terms of the
velocity, the momentum equation relates the velocity to the electric field.
Then from the momentum equation we solve for the velocity in terms of the
electric field to find the susceptibility tensor for each particle species.
Combining this with Maxwell's equations, the result becomes 
\begin{equation}
\left( \delta _{ij}\left( 1-\frac{k^{2}c^{2}}{\omega ^{2}}\right) +\frac{%
k_{i}k_{j}c^{2}}{\omega ^{2}}+\chi _{ij}\right) E^{j}=0  \label{Eq:Matrix-1}
\end{equation}
with $\delta _{ij}$ is the Kronecker delta and the susceptibility tensor is
\begin{equation}
\chi _{ij}=\sum_{s}\left( 
\begin{array}{ccc}
-\frac{\omega _{ps}^{2}(\omega ^{2}-k_{z}^{2}v_{ts}^{2})}{\omega _{d}^{4}} & 
-i\frac{\omega _{ps}^{2}\omega _{cs}(\omega ^{2}-k_{z}^{2}v_{ts}^{2})}{%
\omega \omega _{d}^{4}} & -\frac{\omega _{ps}^{2}k_{\bot }k_{z}v_{ts}^{2}}{%
\omega _{d}^{4}} \\ 
&  &  \\ 
i\frac{\omega _{ps}^{2}\omega _{cs}(\omega ^{2}-k_{z}^{2}v_{ts}^{2})}{\omega
\omega _{d}^{4}} & -\frac{\omega _{ps}^{2}(\omega ^{2}-k^{2}v_{ts}^{2})}{%
\omega _{d}^{4}} & i\frac{\omega _{ps}^{2}\omega _{cs}k_{\bot
}k_{z}v_{ts}^{2}}{\omega \omega _{d}^{4}} \\ 
&  &  \\ 
-\frac{\omega _{ps}^{2}k_{\bot }k_{z}v_{ts}^{2}}{\omega _{d}^{4}} & -i\frac{%
\omega _{ps}^{2}\omega _{cs}k_{\bot }k_{z}v_{ts}^{2}}{\omega \omega _{d}^{4}}
& -\frac{\omega _{ps}^{2}(\omega ^{2}-k_{\bot }^{2}v_{ts}^{2}-\omega
_{cs}^{2})}{\omega _{d}^{4}}
\end{array}
\right)   \label{Eq:suscept}
\end{equation}
where $v_{ts}=\left( k_{B}T/m_{s}\right) ^{1/2}$ is the thermal velocity, $%
\omega _{ps}=(n_{0}q_{s}^{2}/\varepsilon _{0}m_{s})^{1/2}$ the plasma
frequency, $\omega _{cs}=q_{s}B_{0}/m_{s}$ the gyrofrequency and $\omega
_{d}^{4}=\left( \omega ^{2}-k_{\bot }^{2}v_{ts}^{2}\right) \left( \omega
^{2}-k_{z}^{2}v_{ts}^{2}\right) -\omega _{cs}^{2}\left( \omega
^{2}-k_{z}^{2}v_{ts}^{2}\right) -k_{\bot }^{2}k_{z}^{2}v_{ts}^{4}$.

The next aim is to add the electron spin contribution, where the spin
effects due to the ions is neglected due to their small magnetic moment. In general the
spin vector is a dynamical variable whose relation to the EM-field is
complex already in the linear theory. Thus from now on, we limit ourselves to
the case where the dynamics is slow compared to the spin precession period,
which is equivalent to limiting ourselves to frequencies $\omega \ll \left|
\omega _{ce}\right| $ In that case we can write the (electron) spin vector
as 
\begin{equation}
\mathbf{s=-}\frac{\hbar }{2}\widehat{\mathbf{B}}\tanh \left( \frac{\mu
_{B}B_{0}}{k_{B}T}\right)   \label{Eq:spin-vector}
\end{equation}
where the $\tanh $-factor is due to thermodynamic considerations.\footnote{%
Using Fermi-dirac statistics, the slight overweight of particles with spin
orientation in the lower energy state results in a macroscopic spin-vector
proportional to $\tanh \left( \mu _{B}B_{0}/k_{B}T\right) $.} The linearized
magnetization current then becomes 
\begin{equation}
\mathbf{J}_{m}=-\frac{\mu _{B}}{2}\tanh \left( \frac{\mu _{B}B_{0}}{k_{B}T}%
\right) \left( \nabla n_{1}\times \widehat{\mathbf{z}}+n_{0}\nabla \times
\left( \frac{\mathbf{B}_{1}-B_{1z}\widehat{\mathbf{z}}}{B_{0}}\right)
\right)   \label{Eq:magn-current}
\end{equation}
The magnetic field $\mathbf{B}_{1}$ can be expressed in terms of $\mathbf{E}$
from Faradays law, and the density is given in terms of the velocity from
the continuity equation, which is expressed in terms of $\mathbf{E}$ from
the given susceptibility of each species. Thus the magnetization current (%
\ref{Eq:magn-current}) is also expressed in terms of $\mathbf{E}$. However,
before that procedure is implemented, we must also modify the standard
susceptibility Eq. (\ref{Eq:suscept}) to account for the spin dependent force in
the momentum equation. This can be achieved by noting that when solving the
momentum equation for the velocity, the magnetic dipole spin force can
simply be accounted for by including the different components as ''effective
electric fields'' . \ Thus when solving for the electron velocity in terms
of the electric fields, the spin force is included simply by making the
substitutions 
\begin{equation}
\begin{array}{c}
\overline{E}_{jx}=E_{x}+\tanh \left( \frac{\mu _{B}B_{0}}{k_{B}T}\right) 
\frac{i\hbar k_{x}}{m_{e}}B_{1z} \\ 
\overline{E}_{jy}=E_{y} \\ 
E_{z}+\tanh \left( \frac{\mu _{B}B_{0}}{k_{B}T}\right) \frac{i\hbar k_{z}}{%
m_{e}}B_{1z}
\end{array}
\label{Eq:subst}
\end{equation}
Again expressing $B_{1z}$ in terms of $\mathbf{E}$\textbf{\ }through
Faraday's law, these alterations can be expressed as a spin modification of
the free current susceptibility. Thus formally we can write 
\[
j^{i}=j_{free}^{i}+j_{sp}^{i}=\chi _{free}^{ij}E_{j}+\chi _{sp}^{ij}E_{j}
\]
where the direct spin magnetization contained in $\chi _{sp}^{ij}$ can be
determined from (\ref{Eq:magn-current}) (by expressing $\mathbf{B}_{1}$ and $%
n_{1}$ in terms of $\mathbf{E}$), and the free part of the susceptibility
can be divided as 
\[
\chi _{free}^{ij}=\chi _{L}^{ij}+\chi _{md}^{ij}
\]
where $\chi _{L}^{ij}$ is the (free current) susceptibility due to the
Lorentz force given by (\ref{Eq:suscept}), and the contribution from the
magnetic dipole force $\chi _{md}^{ij}$ can be found from (\ref{Eq:suscept})
combined with the substitution in (\ref{Eq:subst}) and the z-component of (%
\ref{Eq:Faraday}). The theory outlined here is straightforward, but results 
in rather cumbersome formulas. To reduce the complexity, and arrive at more transparent expressions, we
introduce the following simplifications:

\begin{enumerate}
\item  The plasma is quasi-neutral, which is a valid approximation provided $%
\omega _{pi}^{2}\gg \omega _{ci}^{2}$

\item  The perpendicular (to the magnetic field) free electron (i.e.
non-spin) part of the current can be approximated by the $\mathbf{E\times B}$%
-drift\textbf{, }$\ $which is valid for frequencies well below the electron
gyro frequency.

\item  The displacement current in Maxwells equations is small. This is
valid when point 1 applies together with $\omega _{pi}^{2}\gg \omega ^{2}$,
and amounts to neglecting the Kronecker delta term in (\ref{Eq:Matrix-1})

\item  Only electron thermal motion is of significance, which is valid if $%
k^{2}v_{ti}^{2}\ll \omega ^{2}$.
\end{enumerate}

The theory outlined above now reduces to Eq. (\ref{Eq:Matrix-1}) with $\chi
_{ij}$ given by 
\begin{equation}
\chi =\left( 
\begin{array}{ccc}
-\frac{\widetilde{\omega }_{pi}^{2}}{\omega ^{2}-\omega _{ci}^{2}} & -i\frac{%
\widetilde{\omega }_{pi}^{2}\omega }{\omega _{ci}(\omega ^{2}-\omega
_{ci}^{2})} & -\frac{\widetilde{\omega }_{pi}^{2}k_{\bot }k_{z}\widetilde{c}%
_{s}^{2}}{\omega _{ci}^{2}} \\ 
&  &  \\ 
i\frac{\widetilde{\omega }_{pi}^{2}\omega }{\omega _{ci}(\omega ^{2}-\omega
_{ci}^{2})} & -\frac{\widetilde{\omega }_{pi}^{2}(\omega ^{2}-k^{2}%
\widetilde{c}_{s}^{2})}{\omega ^{2}-\omega _{ci}^{2}} & i\frac{\widetilde{%
\omega }_{pi}^{2}k_{\bot }k_{z}\widetilde{c}_{s}^{2}}{\omega \omega _{ci}}
\\ 
&  &  \\ 
-\frac{\widetilde{\omega }_{pi}^{2}k_{\bot }k_{z}\widetilde{c}_{s}^{2}}{%
\omega _{ci}^{2}} & -i\frac{\widetilde{\omega }_{pi}^{2}k_{\bot }k_{z}%
\widetilde{c}_{s}^{2}}{\omega \omega _{ci}} & -\frac{\omega _{pi}^{2}}{%
\omega ^{2}}\left( 1-\frac{\omega ^{2}}{k_{z}^{2}c_{s}^{2}}\right) 
\end{array}
\right)   \label{Eq:sus-spec}
\end{equation}
where $\widetilde{\omega }_{pi}=\omega _{pi}(1-\mu _{0}M_{0}/B_{0})^{-1/2}$
is the spin modified ion plasma frequency, $\widetilde{c}_{s}=[c_{s}^{2}-(%
\mu _{B}B_{0}/m_{i})\tanh (\mu _{B}B_{0}/k_{B}T)]^{1/2}$ is the spin
modified ion-acoustic velocity and $c_{s}=v_{ti}(m_{e}/m_{i})^{1/2}$ is the
standard ion-acoustic velocity. Furthermore, the unperturbed magnetization $%
M_{0}$ is given by $M_{0}=n_{0}\mu _{B}\tanh (\mu _{B}B_{0}/k_{B}T)$. We
note that by contrast to the other tensor components, $\chi _{33}$ is not
modified by the spin effects. As a specific example, we can investigate the
magnetohydrodynamic limit $\omega \ll \omega_{ci} $, in which case we get the
dispersion relation 
\begin{equation}
\left( \omega ^{2}-k_{z}^{2}\widetilde{C}_{A}^{2}\right) \left[ \left(
\omega ^{2}-k^{2}\widetilde{C}_{A}^{2}-k_{\perp }^{2}\widetilde{c}%
_{s}^{2}\right) \left( \omega ^{2}-k_{z}^{2}c_{s}^{2}\right) -k_{\perp
}^{2}k_{z}^{2}\widetilde{c}_{s}^{4}\right] =0.  \label{Eq:MHD-DR}
\end{equation}
by putting the determinant of (\ref{Eq:Matrix-1}) to zero. Here $\widetilde{C%
}_{A}=C_{A}(1-\mu _{0}M_{0}/B_{0})^{1/2}$ is the spin-modified Alfv\'{e}n
velocity, and $C_{A}=(B_{0}^{2}/\mu _{0}m_{i}n_{0})^{1/2}$ is the standard
Alfv\'{e}n velocity. Eq. (\ref{Eq:MHD-DR}) agrees with the results of Refs. 
\cite{brodin-marklund2} in the appropriate limiting cases, provided a sign
error is corrected in the last term of their dispersion relation. The first
factor of Eq. (\ref{Eq:MHD-DR}) describes the shear Alfv\'{e}n mode, whereas
the second factor has two roots, describing the fast and slow magnetosonic
modes.

\section{The possibility of a plasma as a metamaterial}

Next we are interested in the possibilities to get metamaterial properties.
As a starting point we note from (\ref{Eq:sus-spec}) and (\ref{Eq:Matrix-1})
that the (relative) dielectric component $\varepsilon _{xx}$ is given by 
\begin{equation}
\varepsilon _{xx}=-\frac{\omega _{pi}^{2}}{\omega ^{2}-\omega _{ci}^{2}}
\label{Eq:exx}
\end{equation}
which become negative for frequencies above the ion cycclotron frequency. 
Here the first Kronecker delta term in (\ref{Eq:Matrix-1}) has been neglected due
to point 3 above. Next we note that the relative magnetic permeability is
given by $\mu _{r}=$ $B_{0}/(B_{0}-\mu _{0}M_{0})$, which causes the
transition from $\omega _{pi}^{2}\longrightarrow \widetilde{\omega }_{pi}^{2}
$ in (\ref{Eq:sus-spec}). In thermodynamic equilibrium, the spin
magnetization $M_{0}$ $=\mu _{B}n_{0}\tanh (\mu _{B}B_{0}/k_{B}T)$ enhances
the external field, which correspond to a paramagnetic situation where $\mu
_{r}>1$. However, let us now assume that we have a laboratory plasma
immersed in an external magnetic field, where the internal spin
magnetization gives a significant contribution to $B_{0}$. Then consider
what happens if we rapidly switch direction of the external field $%
180^{\circ }$, and study the properties before the spin state has time to
reach a new thermodynamic equilibrium state. In particular we are interested
in the case where the external contribution to $B_{0}$ is smaller than the
internal contribution $\mu _{0}M_{0}$, and directed in the opposite
direction. The above linearized theory then still applies, but with the
difference that $\mu _{r}=$ $B_{0}/(B_{0}-\mu _{0}M_{0})$ is now is a
negative quantity. A first observation of the changed properties of this
system is that long wavelength waves (i.e. $kC_{A}<$ $\omega _{ci}$)
described in (\ref{Eq:MHD-DR}) is now unstable, since $\widetilde{C}_{A}^{2}$
changes sign with $\mu _{r}$ and becomes negative for the above scenario. We
shall assume that it is still useful to study the stable waves with shorter
wavelengths, however. In particular this is of interest in case one of the
following conditions apply

\begin{enumerate}
\item  The plasma system is of a rather small size, and the long wavelength
waves ( i.e. $kC_{A}<$ $\omega _{ci}$) are stabilized by inhomogeneities not
included in the model.

\item  The growth rate $\gamma $ of the long wavelength (at most of the
order $\gamma \sim \omega _{ci}$) is slow enough such that there is still
time to study the physics on a time scale much shorter than $\omega
_{ci}^{-1}$

\item  Some dissipative mechanism not included in our model will be
sufficient to stabilize the instability (however, note that the addition of
a finite resistivity will not suffice for this purpose).
\end{enumerate}

As a specific example we study transverse waves with $\omega >\omega _{ci}$
propagating along $B_{0}$ (that is in the opposite direction of $B_{0}-\mu
_{0}M_{0}$). The dispersion relation then reduces to: 
\begin{equation}
\frac{\omega ^{2}\omega _{ci}}{\omega _{ci}\mp \omega }=k^{2}\widetilde{C}%
_{A}^{2}  \label{Eq:DR-par}
\end{equation}
where the $+(-)$ sign corresponds to a right hand (left hand) circular
polarization. As described above, both $\mu _{r}$ and $\widetilde{C}_{A}^{2}$
changes sign with $(B_{0}-\mu _{0}M_{0})$, and thus (\ref{Eq:DR-par})
confirms that waves are unstable in the long wavelength regime. However, for
shorter wavelengths, $\left| k\widetilde{C}_{A}\right| >2\omega _{ci}$, the
waves are stable independently of the sign of $\widetilde{C}_{A}^{2}$. Next
considering this special case, such that $k$ and $\omega $ are real, we find
that the time-averaged Poynting vector $\left\langle \mathbf{S}\right\rangle 
$ is 
\begin{equation}
\left\langle \mathbf{S}\right\rangle =\left\langle \mathbf{E\times H}%
\right\rangle =\left| \mathbf{E}\right| ^{2}\frac{k}{\omega }\frac{%
(B_{0}-\mu _{0}M_{0})}{B_{0}}  \label{Eq:Poynting}
\end{equation}
which apparently changes sign with $B_{0}-\mu _{0}M_{0}$ such that the above
system shows the characteristics of a metamaterial.

\section{Summary and Conclusion}

Previous theories \cite{marklund-brodin,brodin-marklund} on magnetized spin
plasmas have been extended. In particular, the linear theory have been
generalized to cover the frequency range $\omega _{ci}\lessapprox \omega \ll
\omega _{ce}$. The purpose have been to investigate whether it is possible
to produce the characteristics of a metamaterial. It is found that in
principle this is possible by switching the direction of an external
magnetic field $180^{\circ }$, provided the spin magnetization of the plasma
is sufficiently large. However, it is a great challenge to produce the
desired plasma conditions in the laboratory. In particular we need to
combine low temperatures with high densities to obtain a sufficient
magnetization. Ways to reach high plasma densities have been known for a
rather long time, and methods to reach extremely low plasma temperatures has
recently been found \cite{li-etal,fletcher-etal}. On the other hand, the
ultra cold plasmas are still of too low density to be useful for the above
purpose. Laser produced plasmas can reach sufficient densities, and it is
possible that experiments can be designed to keep the temperature
sufficiently low. A challenge with such a setting might be that the plasma
background dynamics is too fast for successful experiments of this kind to
be done. In conclusion, the production of a plasma metamaterial consists of
two challenges. Firstly, to produce a plasma with a sufficient
magnetization, and secondly to master the long wavelength instabilities that
are introduced when the direction of the external field is changed.

\end{document}